\documentclass[aps,preprint,amsmath,amssymb,t1,11pt]{revtex4}
\usepackage{graphicx}
\usepackage{epstopdf}
\usepackage{xcolor}
\usepackage{slashed}
\usepackage{multirow}

\begin{document}
\draft

\title{Probing the electromagnetic properties of the neutrinos at future lepton colliders}

\author{H. Denizli}
\email[]{denizli_h@ibu.edu.tr}
\affiliation{Department of Physics, Bolu Abant Izzet Baysal University, 14280, Bolu T\"{u}rkiye.}

\author{A. Senol}
\email[]{senol_a@ibu.edu.tr}
\affiliation{Department of Physics, Bolu Abant Izzet Baysal University, 14280, Bolu, T\"{u}rkiye.}

\author{M. K{\"o}ksal}
\email[]{mkoksal@cumhuriyet.edu.tr}
\affiliation{Department of Physics, Sivas Cumhuriyet University, 58140, Sivas, T\"{u}rkiye.}

\date{\today}

\begin{abstract}
In this study, we explore the non-standard $\nu\bar{\nu}\gamma \gamma$ couplings parametrized by dimension-seven operators via $e^{+}e^{-} \to \nu\bar{\nu}\gamma$ process  at the FCC-ee/CEPC and $\mu^{+}\mu^{-}\to \nu\bar{\nu}\gamma$ process at the Muon Colliders. For the detailed Monte Carlo simulation, all signal and relevant background events are produced within the framework of Madgraph where non-standard $\nu\bar{\nu}\gamma \gamma$ couplings are implemented. After passing through Pythia for parton showering and hadronization, detector effects are included via tuned corresponding detector cards for each collider in Delphes. Projected sensitivities on $\nu\bar{\nu}\gamma \gamma$ couplings 
are obtained at a 5$\sigma$ confidence level without and with $5\%$ systematic uncertainties for the FCC-ee/CEPC and the Muon Colliders, showcasing the complementarity between lepton colliders.
Our best limit on the anomalous $\nu\bar{\nu}\gamma \gamma$ couplings even with 5\% systematic uncertainties for muon collider with $\sqrt{s}=10$ TeV and $L_{int}=3$ ab$^{-1}$ are found to be thirteen orders of magnitude stronger than the upper bound obtained from rare decay $Z\to\gamma\gamma\nu\bar{\nu}$ analysis using LEP data.
\end{abstract}


\vspace{5mm}

\maketitle


\section{Introduction}

The Standard Model (SM) is a very powerful tool to predict the characteristics, the behavior and the interactions of the elementary particles. The successful validation of the SM of particle physics represents a major milestone in modern physics.  Thus, it is very important to measure particle properties and interactions in the most accurate way possible to better understand the SM.  Since leptons are point-like particles, a lepton-lepton collider as a very clean initial state. Therefore, the envisaged future lepton colliders could provide accurate measurements of SM properties as well as probe clues beyond the SM. There are several proposed designs currently being developed and explored as future lepton colliders. Among these, the Future Circular Collider electron-positron (FCC-ee)\cite{fcc}, the Circular Electron Positron Collider (CEPC) \cite{cepc, cepc2} and Muon Collider come to front \cite{u1,u2}.

The most powerful post-LHC experimental infrastructure FCC has two stage. Starting with electron-positron collider (FCC-ee) \cite{fcc} phase will be followed by proton-proton collider (FCC-hh)\cite{fcc2} in the same tunnel at CERN. There are four proposed stages at different center-of-mass energies for the FCC-ee. The first stage will operate at the $Z$ pole, collecting data for four years and producing approximately 10$^{12}$ $Z$ bosons with an expected data yield of 150 ab$^{-1}$. Next, a two-year run will take place at the $W^{+}W^{-}$ production threshold with an integrated luminosity of 3.8 ab$^{-1}$, 10$^{8}$ or 10$^{12}$ $W^{+}W^{-}$ pairs at a collision energy of $\sqrt{s}=160$ GeV. After this, there will be a three-year run at a collision energy of $\sqrt{s}=240$ GeV to produce the Higgsstrahlung process $e^{+}e^{-}\to ZH$, which will result in one million $ZH$ events with an integrated luminosity of 5 ab$^{-1}$. Finally, a $t\bar{t}$ threshold run will be performed as a multipoint scan around the threshold range of $\sqrt{s}=345-360$ GeV. The expected integrated luminosity for this run is 1.5 ab$^{-1}$, resulting in 10$^{6}$ $t\bar{t}$ events.

One of the other future circular lepton colliders proposed by the Chinese high energy physics community in 2012 is the Circular Electron Positron Collider (CEPC). It is designed to run primarily as a Higgs factory at the center-of-mass energy of 240 GeV with an integrated luminosity up to 5.6 ab$^{-1}$. It will also be operated on the Z-pole, WW threshold scan and at the $t\bar{t}$ threshold with center-of-mass energy of 360 GeV.

Another post-LHC collider option, Muon Collider, was recommended in the 2020 Update of the European Strategy for Particle Physics (ESPPU) for the first time in Europe. The limit of the energy reach for Muon Colliders has not been identified yet. However, ongoing studies focus on 3 TeV and 10 TeV designs with an integrated luminosity goal of up to 10 ab$^{-1}$.

Neutrino oscillations produced by neutrino masses and mixing have been observed in numerous experiments \cite{SNO:2001kpb, LSND:2001aii,KamLAND:2002uet,K2K:2006yov,os1, os2,os3,os4,os5}, indicating that neutrinos have a non-zero mass. This discovery has spurred the interest in the electromagnetic properties of neutrinos, as these properties are directly related to the fundamental principles of particle physics. One important application of neutrino electromagnetic properties is in distinguishing between Dirac and Majorana neutrinos. Dirac neutrinos can have both diagonal and off-diagonal magnetic dipole moments, while only off-diagonal moments are possible for Majorana neutrinos. In the SM extension with massive neutrinos, radiative corrections induce tiny couplings of $\nu\bar{\nu}\gamma$ and $\nu\bar{\nu}\gamma \gamma$ \cite{1,2,3,4,5,alex,alex1}. Although the minimal extension of the SM leads to very small couplings, there are several models beyond the SM that predict relatively large couplings. Therefore, further research into the electromagnetic properties of neutrinos is crucial for advancing our understanding of the fundamental principles of particle physics.

Search for the electromagnetic properties of neutrinos in a model-independent way is valuable since it can help to understand physics beyond the SM as well as contribute to studies in astrophysics and cosmology. 
The $\nu\bar{\nu}\gamma$ interactions have potential  to solve the mysterious solar neutrino puzzle which may be due to a large neutrino magnetic moment \cite{mom} or a resonant spin flip caused by Majorana neutrinos \cite{maj}. Experimental limits on the neutrino magnetic moment obtained from neutrino-electron scattering experiments with reactor and solar neutrinos are currently at the order of $10^{-11}\mu_{B}$ \cite{m1,m2,m3,m4,m5,m6}. However, astrophysical observations provide more stringent constraints. For instance, the energy loss of astrophysical objects gives approximately an order of magnitude more restrictive bounds than those obtained from reactor and solar neutrino probes \cite{m7,m8,m9,m10,m11,m12,Alok:2022pdn}. Additionally, $\nu\bar{\nu}\gamma \gamma$ interactions could have a significant impact on a wide range of low- and high-energy reactions of astrophysical and cosmological interest \cite{cos}. For instance, a high rate of photon annihilation into neutrino pairs may explain the observed cooling of stars through neutrino emission \cite{pon}. Other processes of interest involving $\nu\bar{\nu}\gamma \gamma$ interactions include $\nu \gamma \to \nu \gamma$, $\nu\bar{\nu} \to \gamma \gamma$, and the neutrino double-radiative decay $\nu_{i} \to \nu_{j} \gamma \gamma$. 
\section{Theoritical Framework of non-standard $\nu\bar{\nu}\gamma \gamma$ interactions}
The purpose of our study is to investigate the impact of $\nu\bar{\nu}\gamma \gamma$ coupling, defined by an effective Lagrangian approach, on the process $e^{+}e^{-} \to \nu\bar{\nu}\gamma$ at the FCC-ee/CEPC and the process $\mu^{+}\mu^{-}\to \nu\bar{\nu}\gamma$ at the Muon Colliders. Dimension-seven operators parametrize the $\nu\bar{\nu}\gamma \gamma$ coupling at the lowest dimension. The effective Lagrangian is defined as follows \cite{9,10,11,12,13,14}:
\begin{eqnarray}
\label{eq.2}
\mathcal{L}=\frac{1}{4\Lambda^{3}}\bar{\nu}_{i}(\alpha_{R1}^{ij}P_{R}+\alpha_{L1}^{ij}P_{L})\nu_{j}{\widetilde F}_{\mu\nu}{F}^{\mu\nu}+\frac{1}{4\Lambda^{3}}\bar{\nu}_{i}(\alpha_{R2}^{ij}P_{R}+\alpha_{L2}^{ij}P_{L})\nu_{j}{F}_{\mu\nu}{F}^{\mu\nu}.
\end{eqnarray}
where $\Lambda$ shows the new physics scale, ${\widetilde F}_{\mu\nu}=\frac{1}{2}\epsilon_{\mu\nu\alpha\beta}{F}^{\alpha\beta}$, $P_{R}(L)=\frac{1}{2}(1\pm\gamma_{5})$, $\alpha_{Rk}^{ij}$ and $\alpha_{Lk}^{ij}$ are dimensionless coupling constants. In this work, we focus on the Dirac neutrino scenario and aim to determine model-independent limits on $\nu\bar{\nu}\gamma \gamma$ coupling in the effective Lagrangian.

In addition to $\nu\bar{\nu}\gamma \gamma$ coupling,
$\nu\bar{\nu}Z \gamma$ coupling can contribute to  $e^{+}e^{-} (\mu^{+}\mu^{-})\to \nu\bar{\nu}\gamma$ processes. In Refs. \cite{gul,tev}, $\nu \bar{\nu} Z \gamma$ vertex arises from the dimension-eight operators given by
\begin{equation} 
O^{8}_{1}={\mathrm{i}}
(\phi^{\dagger} \phi)\bar{\ell}^{a}_{L}
\tau^{i} \gamma^{\mu}
D^{\nu}\ell^{a}_{L}W^{i}_{ \mu \nu},
\label{eiop1} 
\end{equation}
\begin{equation} 
O^{8}_{2}= {\mathrm{i}}
(\phi^{\dagger} \phi)\bar{\ell}^{a}_{L}
\gamma^{\mu} D^{\nu} \ell^{a}_{L} B_{\mu \nu},
\label{eiop2} 
\end{equation}
\begin{equation}
O^{8}_{3}= {\mathrm{i}} (\phi^{\dagger}
D^{\mu} \phi)\bar{\ell}^{a}_{L} \gamma^{\nu}
\tau^{i} \ell^{a}_{L}W^{i}_{\mu \nu},
\label{eiop3}
\end{equation}
\begin{equation}
O^{8}_{4}= {\mathrm{i}} (\phi^{\dagger} D ^{\mu}
\phi)\bar{\ell}^{a}_{L}
\gamma^{\nu}
\ell^{a}_{L} B_{\mu \nu}. \label{eiop4}
\end{equation}
Here, $W^{i}_{\mu \nu}$ and $B_{\mu \nu}$ show $\mathrm{SU(2)_{L}}$ and $ \mathrm{U(1)_{Y}}$
tensor field strength tensors 
respectively, as well as the  $\ell^{a}_{L}$ represents $\mathrm{SU(2)_{L}}$ left-handed lepton
doublet, $\tilde{\phi}=\mathrm{i} \tau^{2} \phi^{\ast}$ is the Higgs field, $\tau_{i}$ are the Pauli matrices and $D_{\mu}$ is covariant derivative. 

The Lagrangian given in Eq.(\ref{eq.2}) represents the most general dimension-seven Lagrangian that defines the $\nu\bar{\nu}\gamma \gamma$ coupling. On the other hand, $\nu \bar{\nu} Z \gamma$ coupling arising from dimension-eight operators is subject to suppression by an additional power of the high inverse new physics scale $\Lambda$. For this reason, one can disregard their influence in our investigation.

The upper limit on the branching ratio of $Z \to \nu\bar{\nu}\gamma \gamma$ decay from the LEP data was used to derive an upper bound on the $\nu\bar{\nu}\gamma \gamma$ coupling as in Ref. \cite{9}:
\begin{eqnarray}
\label{eq.1}
[\frac{1 GeV}{\Lambda}]^{6}\sum_{i,j,k}(|\alpha_{Rk}^{ij}|^{2}+\alpha_{Lk}^{ij}|^{2})\leq 2.85\times10^{-9}.
\end{eqnarray}

An study of the Primakoff effect on the conversion of $\nu_{\mu}N \to \nu_{\nu}N$ in the external Coulomb field of nucleus $N$ has yielded a bound that is two orders of magnitude more stringent than that obtained from the $Z \to \nu\bar{\nu}\gamma \gamma$ decay \cite{10}. Experimental bounds on the lifetime of the neutrino double radiative decay ($\nu_j\to\nu_i\gamma\gamma$), given in Eq.(\ref{eq.3}), can be used to establish a connection with a typical neutrino mass. With the effective Lagrangian given in Eq.(\ref{eq.2}), the decay width of $\nu_j\to\nu_i\gamma\gamma$ without the mass of $\nu_i$ is obtained as follows \cite{9}:
\begin{equation}
\label{eq.3}
\Gamma_{\nu_j\to\nu_i\gamma\gamma}=1.59\times10^{-9}(|\alpha_{Rk}^{ij}|^{2}+|\alpha_{Lk}^{ij}|^{2})\Big[\frac{1 GeV}{\Lambda}\Big]^6\times\Big[\frac{m_{\nu_j}}{1MeV}\Big]^7 s^{-1}.
\end{equation}

A lot of phenomenological studies beyond the SM were investigated $\nu\bar{\nu}\gamma\gamma$ couplings at $pp$ and $e^{+}e^{-}$ colliders. The limits on $\alpha^2$ by Ref. \cite{7} were obtained through exclusive processes, such as $pp\rightarrow p\gamma^{}\gamma^{}p\rightarrow p\nu\bar{\nu}p$ with a limit of $10^{-16}$ and $pp\rightarrow p\gamma^{}\gamma^{}p\rightarrow p\nu\bar{\nu}Zp$ with a limit of $10^{-17}$ at LHC energies. Ref. \cite{8} obtained sensitivities on $\nu\bar{\nu}\gamma\gamma$ couplings ($\alpha_1^2$ and $\alpha_2^2$) at the order of $10^{-17}$ and $10^{-18}$ via $pp\rightarrow p\gamma p\rightarrow p\nu\bar{\nu}q X$ at the LHC. 
These couplings parametrized with the non-standard dimension-seven operators defined by the effective Lagrangian framework are investigated through the process $pp\rightarrow \nu\bar{\nu}\gamma$ at the HL-LHC and the FCC-hh in Ref. \cite{mur}. They find that the best obtained limits on the anomalous couplings are at the order of $10^{-20}$.
In addition, $e^{+}e^{-}$ linear colliders, including their $e\gamma$ and $\gamma\gamma$ operating modes, have been explored through processes such as $e\gamma\rightarrow\nu\bar{\nu}e$ and $\gamma\gamma\rightarrow\nu\bar{\nu}$ at the CLIC \cite{6}. Ref. \cite{6} has shown that the neutrino-two photon couplings improve the sensitivity limits by up to a factor of $10^{16}$ with respect to the LEP limits. In Refs. \cite{7,8, mur, 6}, limits on the anomalous coupling were obtained with $\Lambda=1$ GeV.

\section{Signal and background analysis for the future lepton colliders}

\begin{figure}[ht]
\includegraphics[scale=0.9]{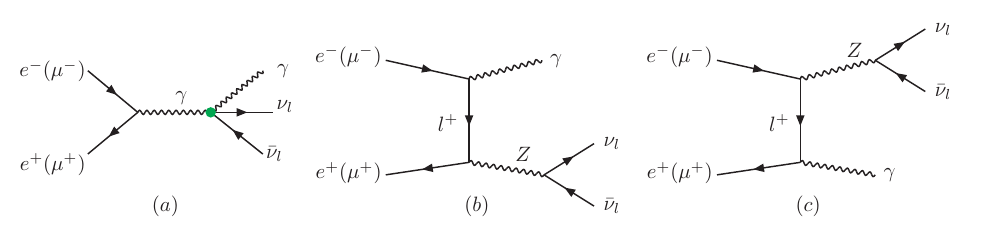}\\\includegraphics[scale=1.0]{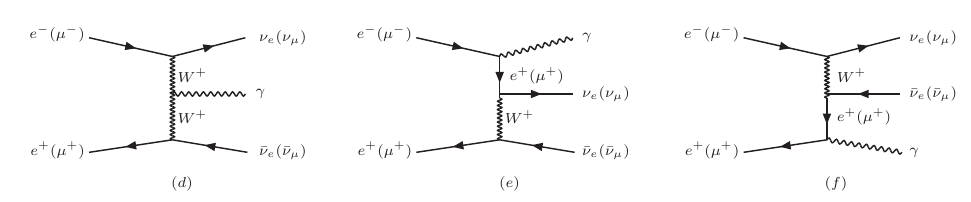}
\caption{The Feynman diagrams contributing $\nu\bar{\nu}\gamma$ production at the future lepton colliders (a) through anomalous $\nu\bar{\nu}\gamma \gamma$ with marked green $(b)$, $(c)$, $(d)$, $(e)$ and $(f)$ as well as SM contributions.}
\label{Fig.1}
\end{figure}

A comprehensive analysis of the $\nu\bar{\nu}\gamma \gamma$ coupling is conducted through the process $e^{+}e^{-}\to \nu\bar{\nu}\gamma$ at the FCC-ee with $\sqrt s=$240, 365 GeV, the CEPC with $\sqrt s=$240 GeV and $\mu^{+}\mu^{-}\to \nu\bar{\nu}\gamma$ at the Muon collider with $\sqrt s=$3, 10 TeV, respectively. 
In the presence of the effective interactions, the processes $e^{+}e^{-} (\mu^{+}\mu^{-}) \to \nu\bar{\nu}\gamma$ are defined by three tree-level Feynman diagrams given in Fig. \ref{Fig.1}. The first diagram corresponds to the anomalous couplings related to new physics (where $l=e,\mu,\tau$), while the remaining diagrams represent the SM contributions. If the flavors of final state neutrinos and initial state charged leptons are the same, the SM background contributions are represented by diagrams $(b)$, $(c)$, $(d)$, $(e)$, $(f)$.  
Therefore, we have calculated the analytical expressions for the polarization summed amplitude square of the processes $e^{+}e^{-} (\mu^{+}\mu^{-}) \to \nu\bar{\nu}\gamma$, which are presented below:

\begin{eqnarray}
\langle|M|^2\rangle&=&\sum_{i,j}\langle|M_a|^2\rangle+\langle|M_b+M_c+M_d+M_e+M_f|^2\rangle.
\end{eqnarray}
As mentioned above, while $M_a$ represents the contribution when the effective interaction $\nu\bar{\nu}\gamma \gamma$ is considered, the contributions $M_b$, $M_c$, $M_d$, $M_e$, $M_f$ correspond to the SM. Since the momentum dependence for both $\alpha_{L1(R1)}^{ij}$ and $\alpha_{L2(R2)}^{ij}$ couplings from the new physics contribution in the amplitude square is the same, we focus on analyze the $\alpha_{L1(R1)}^{ij}$ couplings. 
\begin{figure}[h]
\includegraphics[scale=0.64]{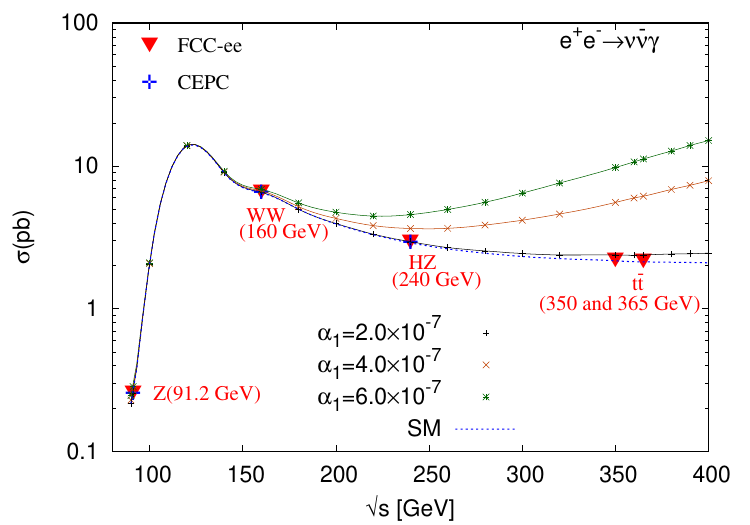}
\includegraphics[scale=0.64]{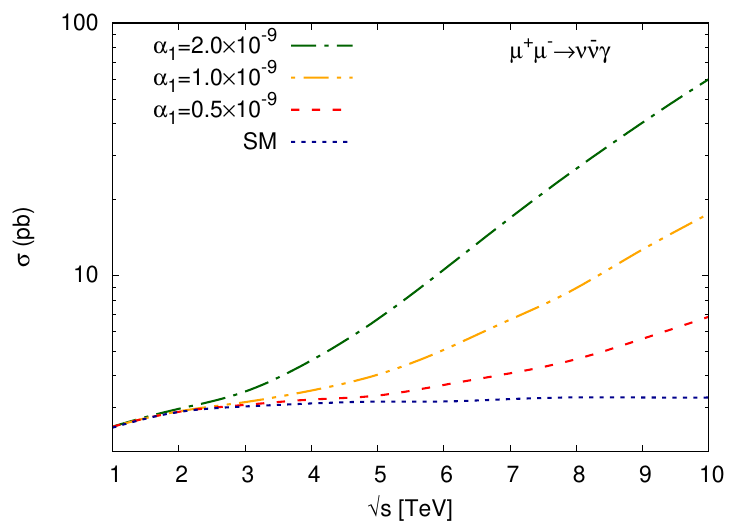}
\caption{The total cross-sections for the process $e^{+}e^{-} (\mu^{+}\mu^{-}) \to \nu\bar{\nu}\gamma$ with respect to center-of-mass energy for some
$\alpha_1$ couplings at the FCC-ee/CEPC on the left panel and at the Muon Colliders on the right panel.}
\label{Fig.2}
\end{figure}
\begin{figure}[ht]
\includegraphics[scale=0.6]{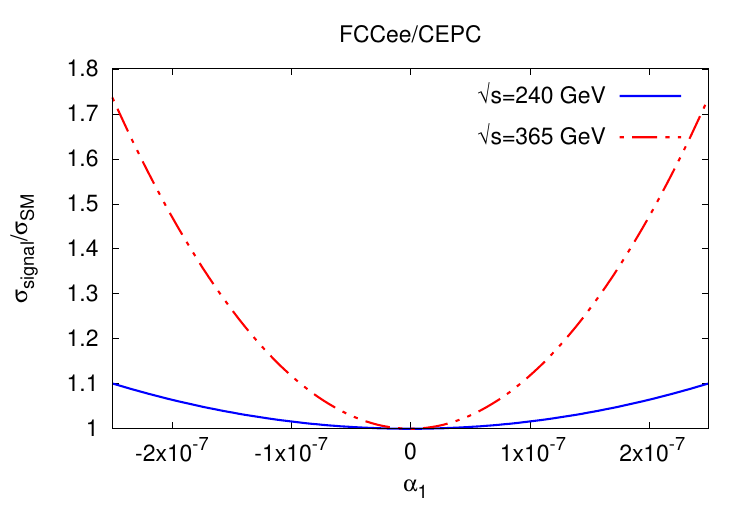}
\includegraphics[scale=0.6]{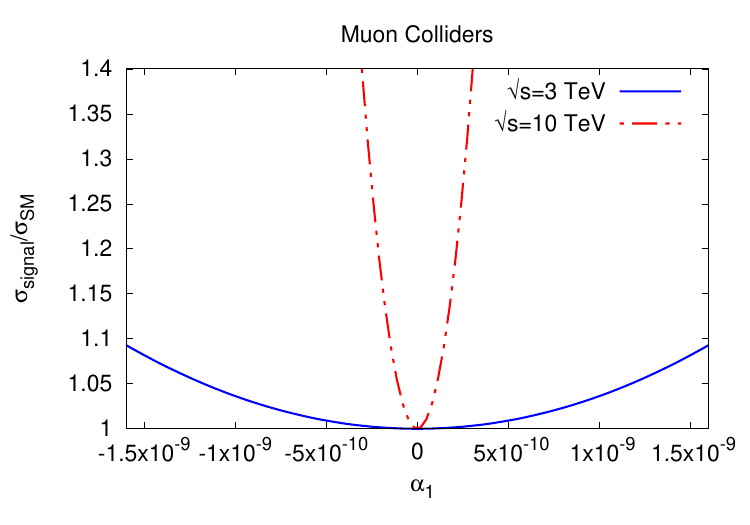}
\caption{The signal-to-SM ratio of total cross-sections for the process $e^{+}e^{-} (\mu^{+}\mu^{-}) \to \nu\bar{\nu}\gamma$ as a function of the anomalous
$\alpha_1$ coupling at the FCC-ee/CEPC on the left panel and at the Muon Colliders on the right panel.}
\label{Fig.3}
\end{figure}

To simulate the signal and background events for the process $e^{+}e^{-} (\mu^{+}\mu^{-}) \to \nu\bar{\nu}\gamma$, we inserted  the {\sc Universal FeynRules Output} (UFO) \cite{ufo} module generated by the {\sc FeynRules} package, where the effective Lagrangian given in Eq.(\ref{eq.2}) is implemented, into {\sc MadGraph5\_aMC$@$NLO v3\_1\_1} \cite{Ball:2012cx}. The total cross-sections for the process $e^{+}e^{-} (\mu^{+}\mu^{-}) \to \nu\bar{\nu}\gamma$ with respect to center-of-mass energy at the FCC-ee/CEPC (Muon Colliders) are shown for some benchmark values of $\alpha_1$ coupling in Fig. \ref{Fig.2} where $\alpha_1$ is defined as $\alpha_1^2=\sum_{i,j}\left[|\alpha^{ij}_{R1}|^2+|\alpha^{ij}_{L1}|^2\right]$. The cross-sections were calculated at leading order with generator level cuts $p^{\gamma}_T>10$ GeV and $|\eta^{\gamma}|<2.5$. In the left panel of Fig. \ref{Fig.2}, several interaction points of the FCC-ee/CEPC: the $Z$ pole, the $WW$ threshold, Higgs factory and the $t\bar t$ threshold are marked to show considered center-of-mass energy of the future colliders. The cross-section values for the Muon Colliders are higher than those for the FCC-ee and CEPC options, as expected due to the higher center-of-mass energy. As seen from the figure, the separability of the signal from the SM at the cross-section level appears after 240 GeV for benchmark values of the new physics parameter $\alpha_1 \approx 10^{-7}$ at the center-of-mass energy region of the FCC-ee/CEPC colliders while 3 TeV for $\alpha_1 \approx 10^{-9}$ at the center-of-mass energy region of the Muon Colliders options. Therefore, we focus on the center-of-mass energies and luminosities of the FCC-ee/CEPC and the Muon Colliders given in Table I for our study.
\begin{table}[htb!]
\caption{The center-of-mass energies and integrated luminosities used for the FCC-ee/CEPC and the Muon Colliders in this study.}
\begin{ruledtabular}
\begin{tabular}{lcccc}
Collider & Center-of-mass energy & Integrated luminosity \\\hline
The FCC-ee / CEPC &240 GeV & 5 ab$^{-1}$ \\
The FCC-ee &365 GeV & 1.5 ab$^{-1}$\\
The Muon Collider &3 /10 TeV & 1 / 3 ab$^{-1}$
\end{tabular}
\end{ruledtabular}
\end{table}
The signal-to-SM ratio of total cross-sections for the process $e^{+}e^{-} (\mu^{+}\mu^{-}) \to \nu\bar{\nu}\gamma$ as a function of the anomalous $\alpha_1$ coupling at the FCC-ee/CEPC on the left panel while the Muon Colliders on the right panel are presented in Fig. \ref{Fig.3}. It is also seen that the deviation from the SM cross-section value begins around $\alpha_1 \approx 10^{-7}$ for the $e^{+}e^{-}$ colliders and $\alpha_1 \approx 10^{-9}$ for the $\mu^{+}\mu^{-}$ colliders. Considering experimental distinguishability on the cross-section due to systematic errors, we have chosen the benchmark values of $\alpha_1$ that cover the cross sections that do differ from SM by 10\% for detailed analysis.

We generate 600k event samples  for the detailed Monte-Carlo analysis of kinematical distributions of all relevant backgrounds as well as signals with seven benchmark values of $\alpha_{L1}^{ij}=\alpha_{R1}^{ij}$ coupling ranging between $5\times10^{-8}$ ($3\times10^{-12}$) and $5\times10^{-7}$($3\times10^{-9}$)
for the FCC-ee/CEPC (Muon Colliders). Parton showering and hadronization are carried out using the {\sc Pythia 8.2} package \cite{pyt}. Subsequently, the events are interfaced with {\sc Delphes 3.4.2} \cite{del} software to model the response of the corresponding detector in the form of resolution functions and efficiencies for the FCC-ee/CEPC and the Muon Colliders. Latest Innovative Detector for electron–positron Accelerators (IDEA) card (delphes\_card\_IDEA.tcl) is chosen for the FCC-ee/CEPC detector concept while the modified detector card based on the performance of the FCC-hh and the CLIC is used for the Muon colliders (delphes\_card\_MuonColliderDet.tcl, respectively). Jets are reconstructed by using clustered energy deposits with FastJet 3.3.2 \cite{Cacciari:2011ma} using the anti-kt algorithm \cite{Cacciari:2008gp}.

Since we investigate the $\nu\bar{\nu}\gamma \gamma$ coupling through the signal process $e^{+}e^{-} (\mu^{+}\mu^{-}) \to \nu\bar{\nu}\gamma$ at the future lepton colliders, a photon and missing energy constitute the final state topology of the signal process. For this reason, we consider the relevant SM background processes that represent the same or similar final state topology below:
\begin{eqnarray}
\label{eq.back}
\nu\nu\gamma&:& e^{+}e^{-} (\mu^{+}\mu^{-}) \to \nu\bar{\nu}\gamma \\
 WW\gamma&:& e^{+}e^{-} (\mu^{+}\mu^{-}) \to WW\gamma ,
\\
Z(qq)\gamma&:& e^{+}e^{-} (\mu^{+}\mu^{-}) \to Z\gamma \,\, \to q\bar q\gamma,
\\
Z(ll)\gamma &:& e^{+}e^{-} (\mu^{+}\mu^{-}) \to Z\gamma \,\, \to l^+l^-\gamma.
\end{eqnarray}

The $\nu\bar{\nu}\gamma$ background originates from all except the first diagram as depicted in Fig.\ref{Fig.1}. The $WW\gamma$ channel is included as a background due to the presence of a dilepton and a neutrino pair resulting from the leptonic decay of $W$ bosons.
The $Z(ll) \gamma$ and $Z(qq) \gamma$ are assumed to be other relevant backgrounds if leptons or quarks coming from the decaying of the $Z$ boson cannot be detected by the detector. 
\section{Results of the analysis}

In this section, in order to achieve high signal efficiency and effective rejection of the relevant SM backgrounds, we focus on obtaining a set of kinematic cuts for particles in the final state of the examined processes $e^{+}e^{-} (\mu^{+}\mu^{-}) \to \nu\bar{\nu}\gamma$.
\subsection{$e^+e^-$ colliders}
We pre-select events based on the final state topology of the signal process (requiring at least one photon and no leptons) and minimum detector requirements (the transverse momentum, Missing Transverse energy and pseudo-rapidity of the photon as $p^{\gamma}_T>10$ GeV, $\slashed{E}_T>$ 10 GeV and $|\eta^{\gamma}|<2.5$, respectively.)
Additionally, we consider the mass of the system recoiling against the photon (photon recoil mass) as the most useful observable in our analysis for single photon events, it is defined as follows:
\begin{equation}
M_{recoil}=\sqrt{s-2\sqrt{s}E_{\gamma}} 
\end{equation}
where $\sqrt{s}$ is the center-of-mass energy and $E_{\gamma}$ is the photon energy.
\begin{figure}[!htb]
\includegraphics[scale=0.4]{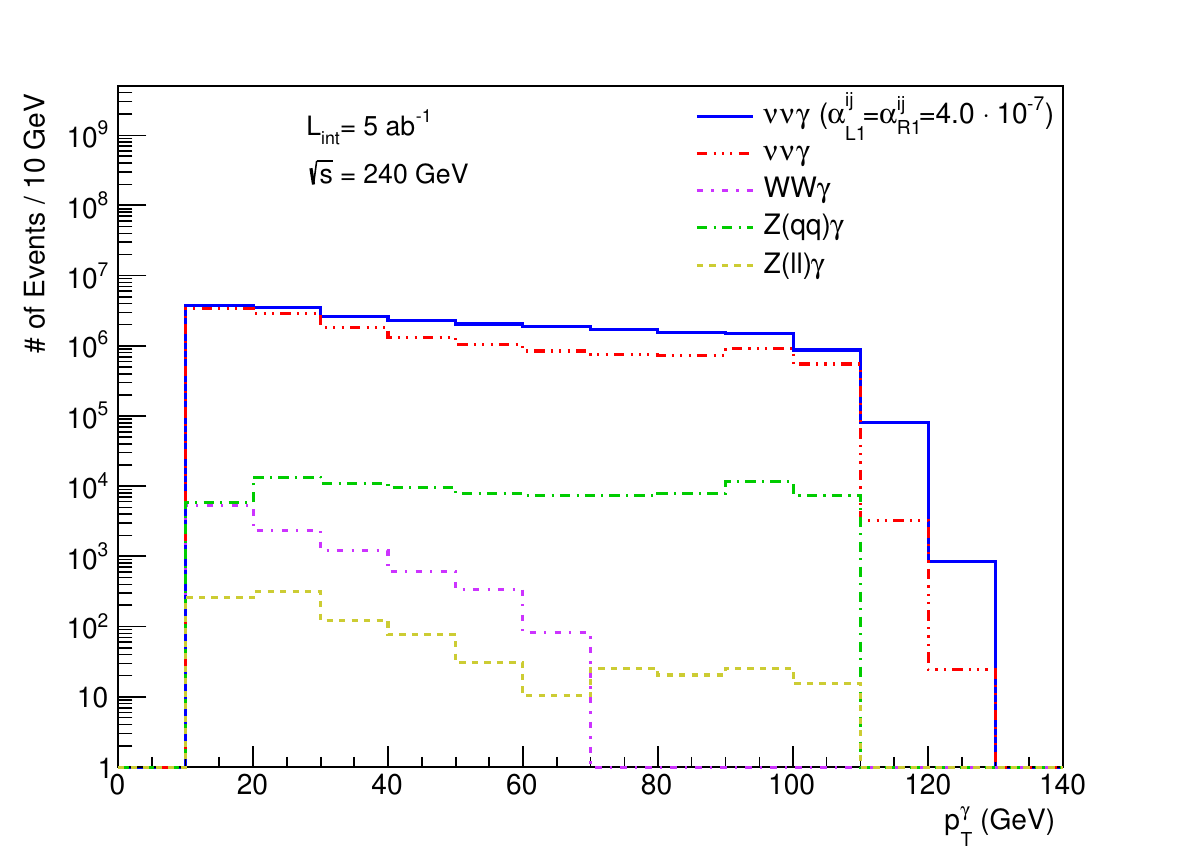}
\includegraphics[scale=0.4]{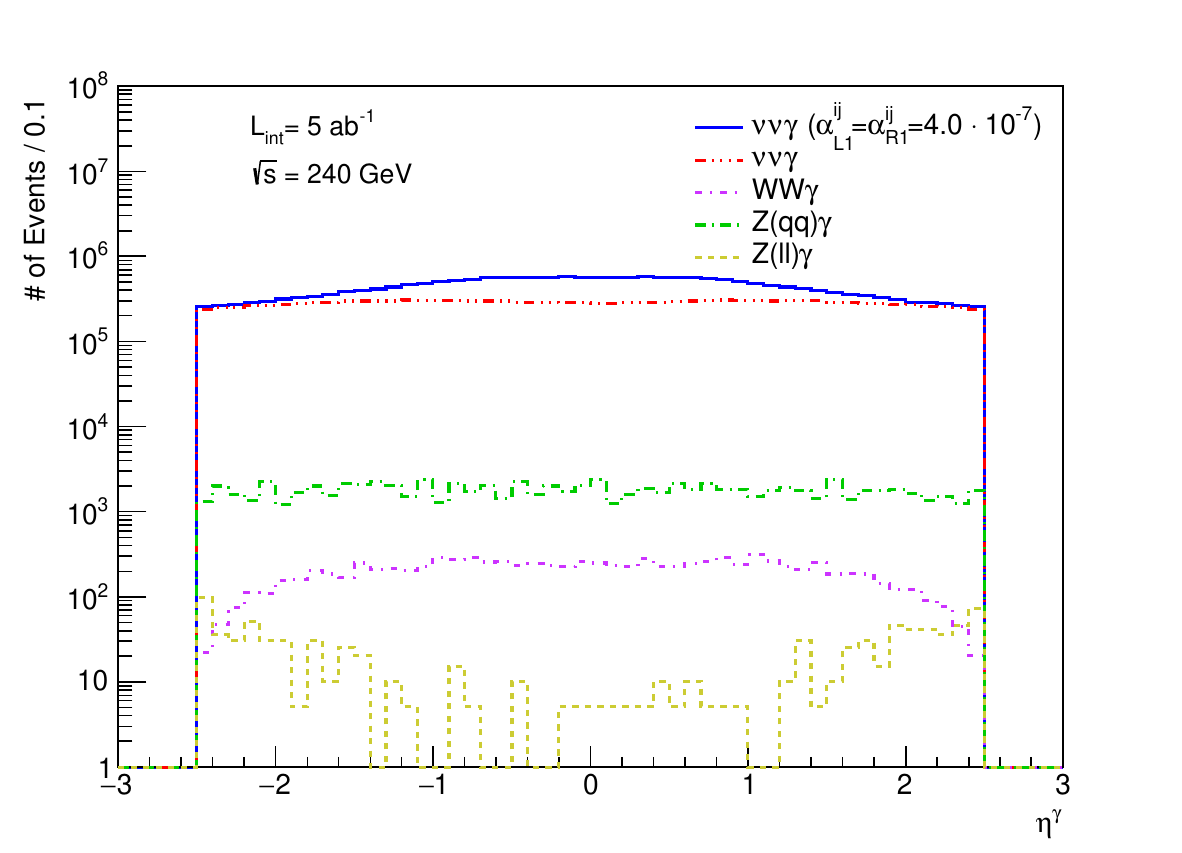}
\includegraphics[scale=0.4]{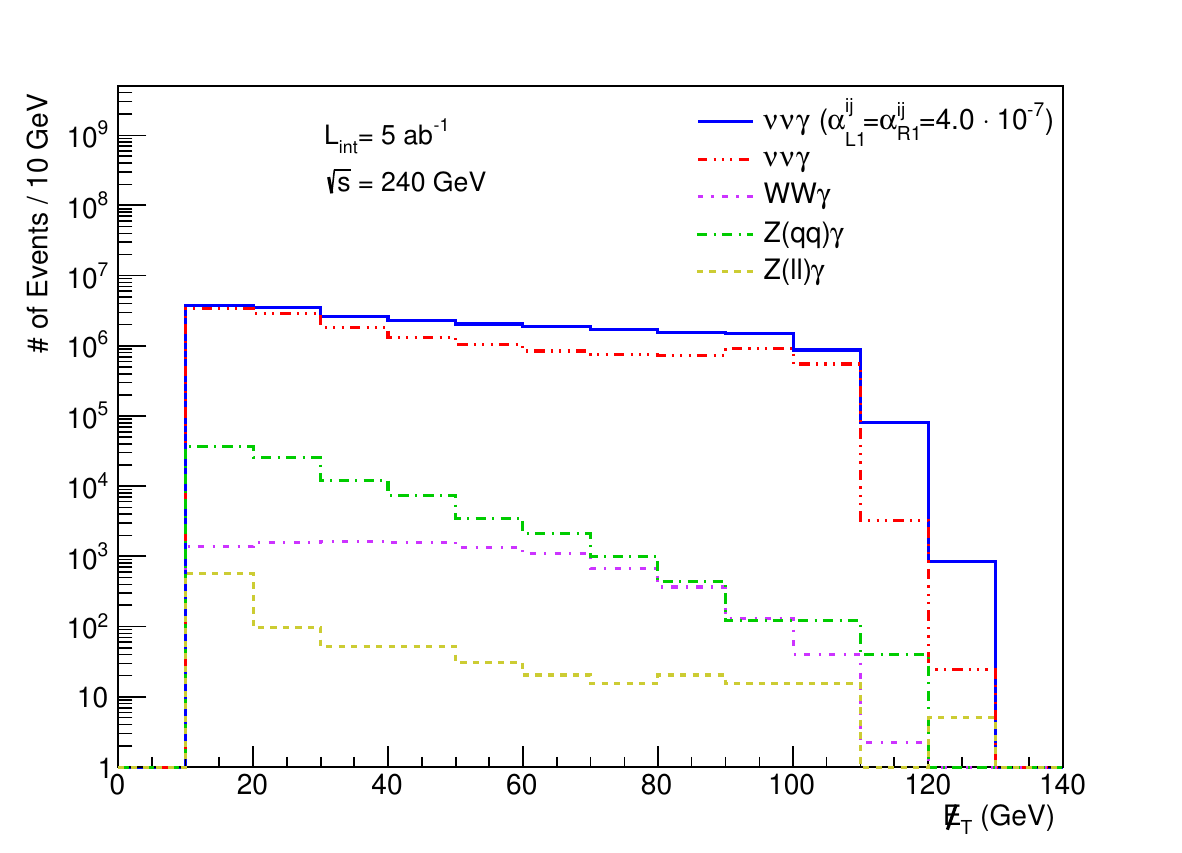}
\includegraphics[scale=0.4]{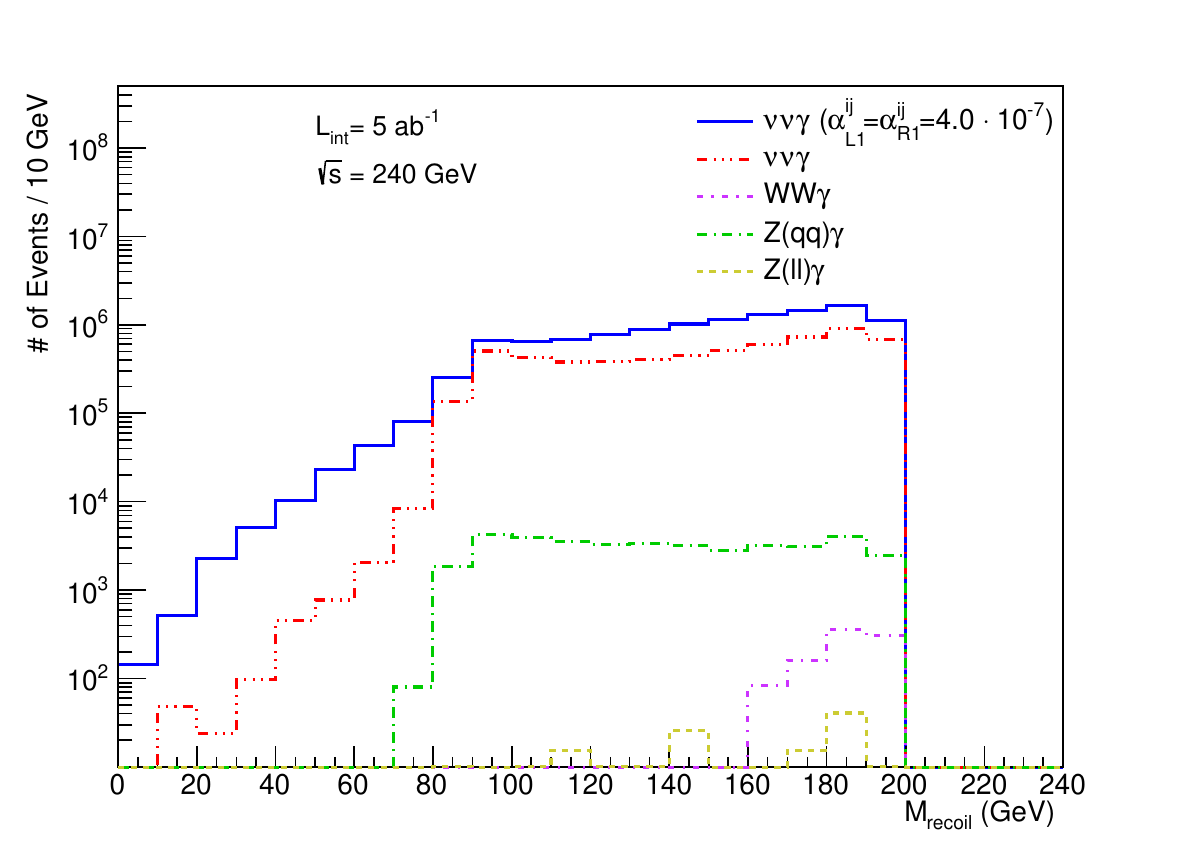}
\caption{The normalized distribution of $p^{\gamma}_T$, $\eta^{\gamma}$, $\slashed{E}_T$ and
$M_{recoil}$ for the $e^{+}e^{-} \to \nu\bar{\nu}\gamma$ signal and relevant SM backgrounds at the FCC-ee/CEPC with $\sqrt{s}=240$ GeV and $L_{int}=5$ ab$^{-1}$.}
\label{Fig.fcc_kinematics}
\end{figure}

In Fig. \ref{Fig.fcc_kinematics}, the normalized distributions of the photon $p^{\gamma}_T$ and $\eta^{\gamma}$ in the first row as well as $\slashed{E}_T$ and
$M_{recoil}$ in the second row are shown for the signal with $\alpha_{L1}^{ij}=\alpha_{R1}^{ij}=4\times 10^{-7}$ (selected for demonstrating purpose) and relevant SM background processes at $\sqrt{s}$= 240 GeV for the FCC-ee/CEPC after the pre-selection cut. 
Similar distributions are also obtained for $\sqrt{s}$= 365 GeV option of the FCC-ee/CEPC. Additional cuts on $\slashed{E}_T$ and $p^{\gamma}_T$ as 20 GeV and 40 GeV have been foreseen after analyzing distributions for all signal events produced ranging between $\alpha_{L1}^{ij}=\alpha_{R1}^{ij}\approx 10^{-7}$ and $\alpha_{L1}^{ij}=\alpha_{R1}^{ij}\approx10^{-8}$ to distinguish the model under study from the relevant SM backgrounds for both energy options. Because we are interested in $\nu\nu\gamma\gamma$ vertex, the photon recoil mass is required to satisfy $M_{recoil} < 70$ GeV and  $M_{recoil} > 110$ GeV so that it is not consistent with that of a $Z$ boson peak. Therefore, we finally impose $M_{recoil} > 110$ GeV in our analysis. The summary of the cuts used in our analysis is given in Table \ref{tab2}.
\begin{figure}[!htb]
\includegraphics[scale=0.4]{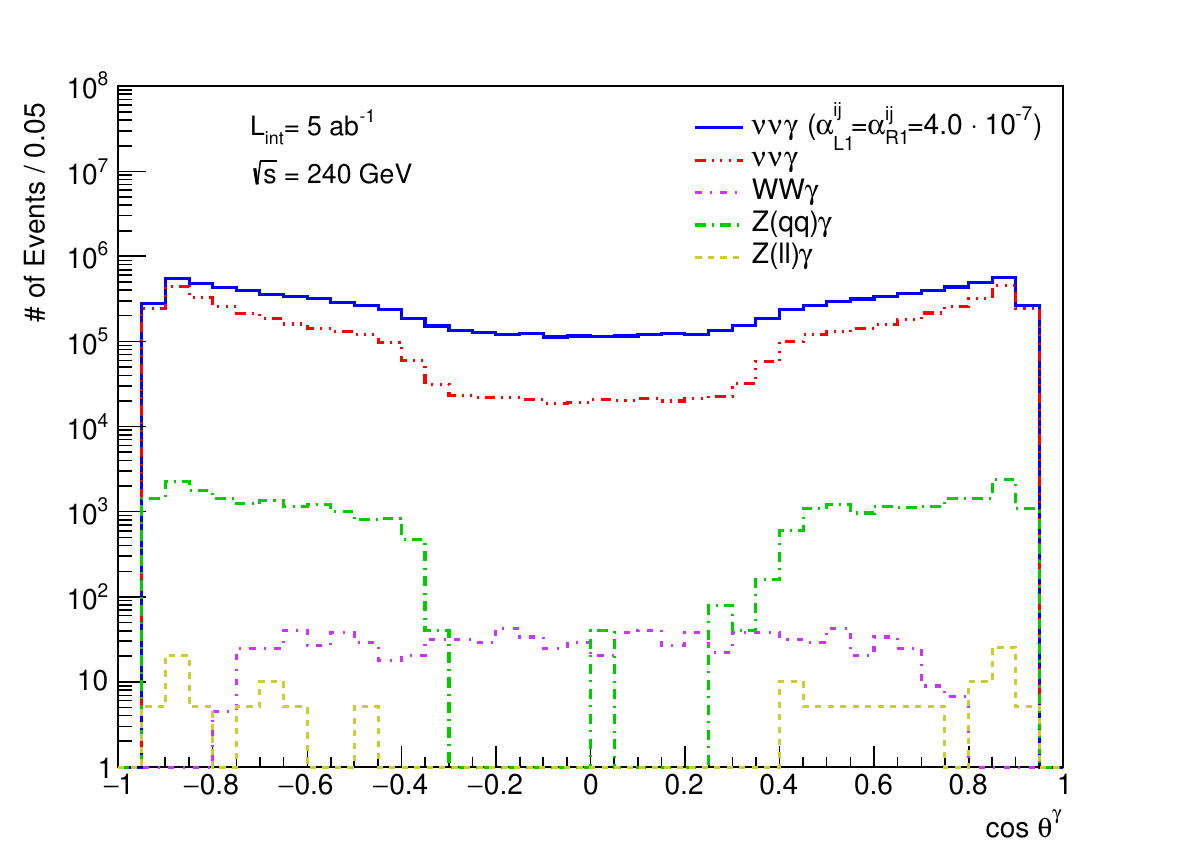}
\includegraphics[scale=0.4]{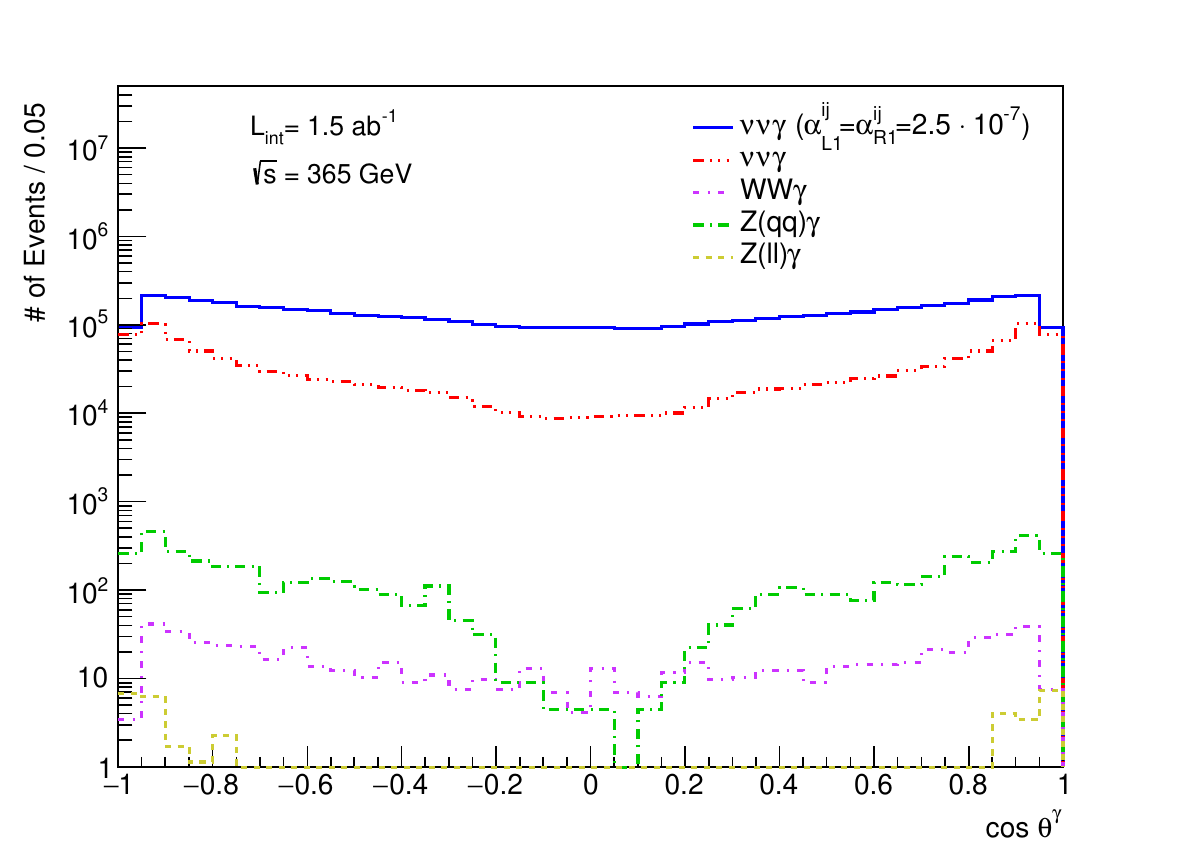}
\caption{The normalized distributions of the polar angle of the photon in the final states for the $e^{+}e^{-} \to \nu\bar{\nu}\gamma$ signal and relevant SM backgrounds at the FCC-ee/CEPC with $\sqrt{s}=240$ GeV (left panel) and at the FCC-ee with $\sqrt{s}=365$ GeV (right panel), respectively.}
\label{Fig.fcc_costheta}
\end{figure}
Since the produced neutrinos are undetectable, such events would lead to the single photon and missing energy signature. Therefore, knowing the transverse momentum as well as the angular distribution of the emitted photon with high precision allows a clear separation of the signal from the relevant SM backgrounds. Fig. \ref{Fig.fcc_costheta} shows the normalized distributions of  $\cos\theta^{\gamma}$, where $\theta^{\gamma}$ is the angle of the photon with respect to the beam direction, for signal and relevant SM backgrounds at $\sqrt{s}$= 240 GeV and $L_{int}=5$ ab$^{-1}$ (on the left panel) and $\sqrt{s}$= 365 GeV and $L_{int}=1.5$ ab$^{-1}$ (on the right panel) for the FCC-ee/CEPC options. As seen from both figures, the shape of the signal is different from the SM backgrounds.
\begin{table}[h]
\caption{Event selection criteria and applied kinematic cuts used for the analysis of the process $e^{+}e^{-} \to \nu\bar{\nu}\gamma$ at $\sqrt s=$240 and 365 GeV options of the FCC-ee/CEPC.}
\label{tab2}
\begin{tabular}{p{3cm}p{13cm}}
\hline
\hline
Cuts & Definitions \\
\hline
Pre-selection &  $N_{\gamma} = 1$ $N_{\ell\,(e,\,\mu)} > 0$, $|\eta^{\gamma}| < 2.5$, $p_T^{\gamma} > 10$ GeV and $E_T > 10$ GeV \\
Cut-1 & Transverse momentum of the photon: $p_T^{\gamma} > 40$ GeV\\
Cut-2 & Missing energy transverse: $\slashed{E}_T > 20$ GeV\\
Cut-3 & Photon recoil mass: $M_{recoil}> 110$ GeV\\
\hline
\hline
\end{tabular}
\end{table}
\subsection{$\mu^+\mu^-$ colliders}

We also analyzed the kinematic features of final state particles for $\mu^{+}\mu^{-} \to \nu\bar{\nu}\gamma$  signal processes and relevant SM backgrounds at 3 and 10 TeV center-of-mass energy options of the Muon Colliders. Before further analysis of the $\mu^{+}\mu^{-} \to \nu\bar{\nu}\gamma$ signal process and the relevant SM backgrounds, the same pre-selection cuts (described in Table \ref{tab2}) used for the $e^{+}e^{-} \to \nu\bar{\nu}\gamma$ analysis are applied.
\begin{figure}[!htb]
\includegraphics[scale=0.4]{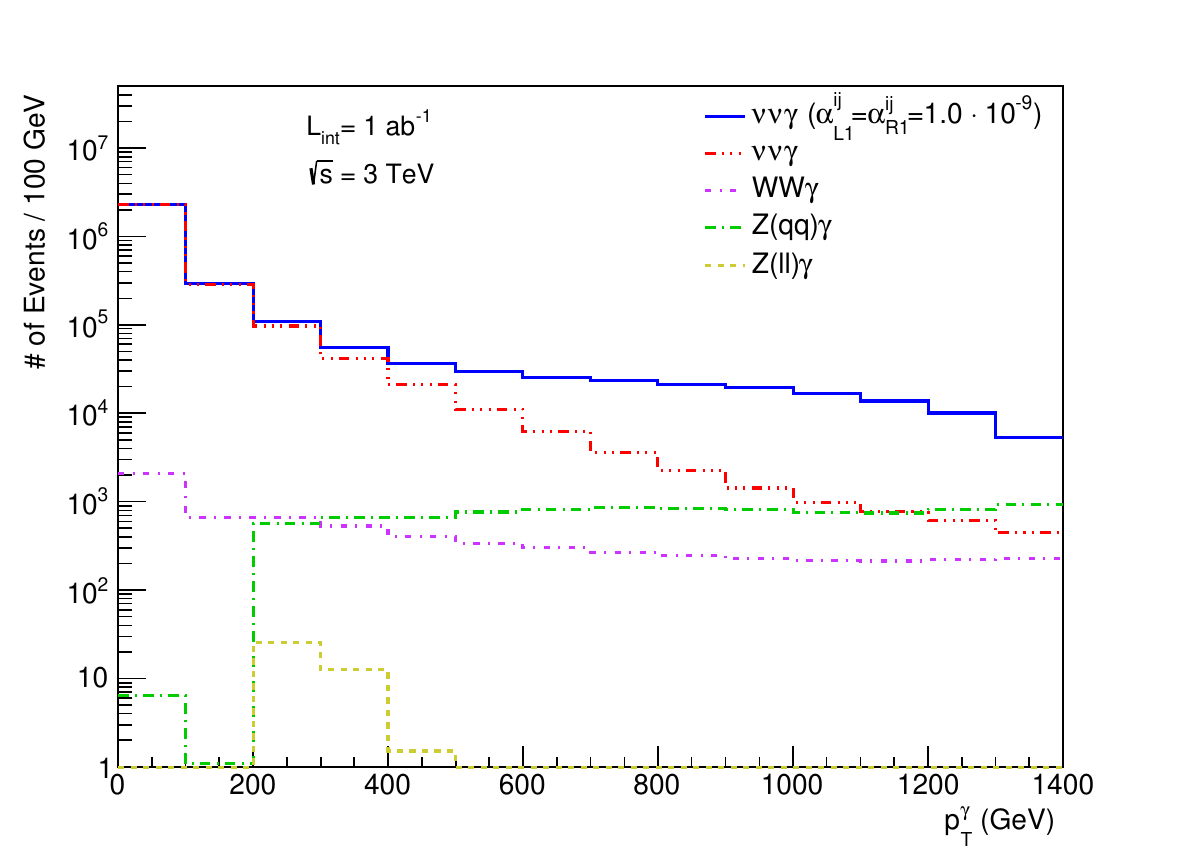}
\includegraphics[scale=0.4]{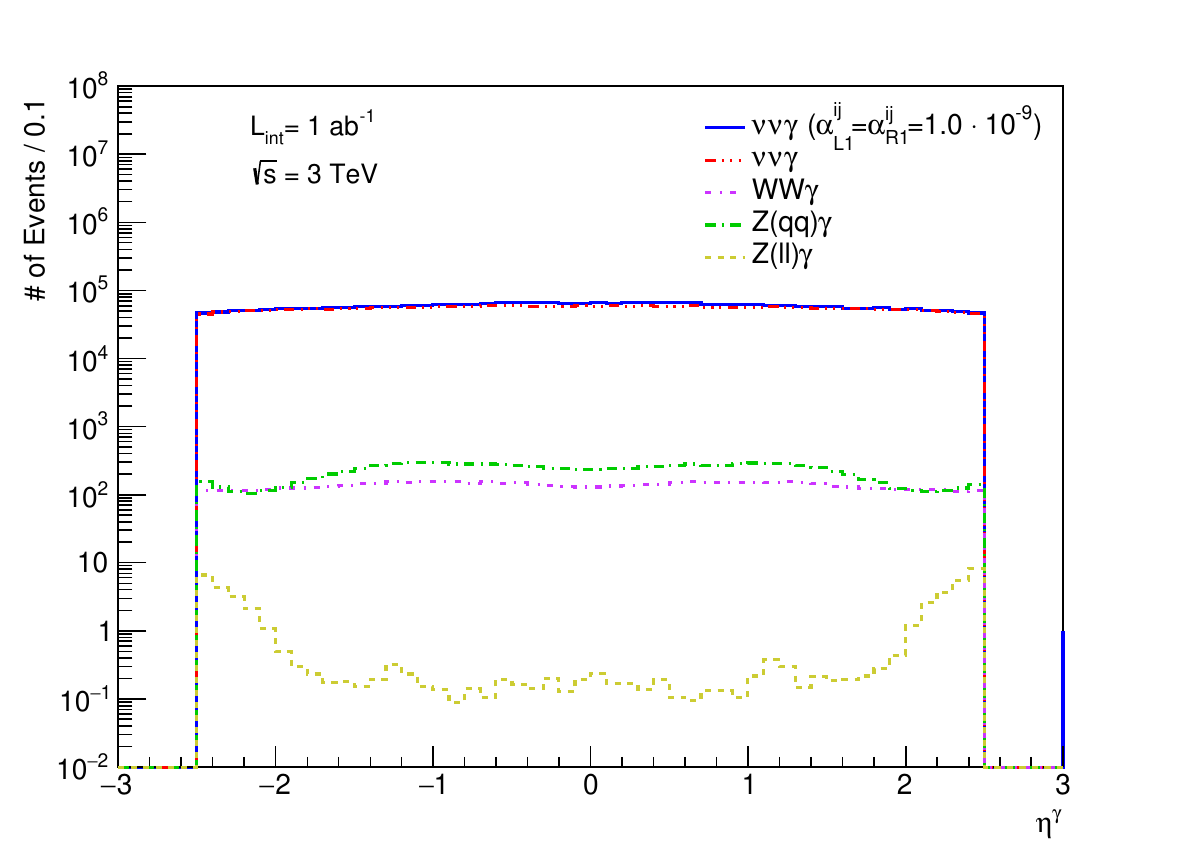}
\includegraphics[scale=0.4]{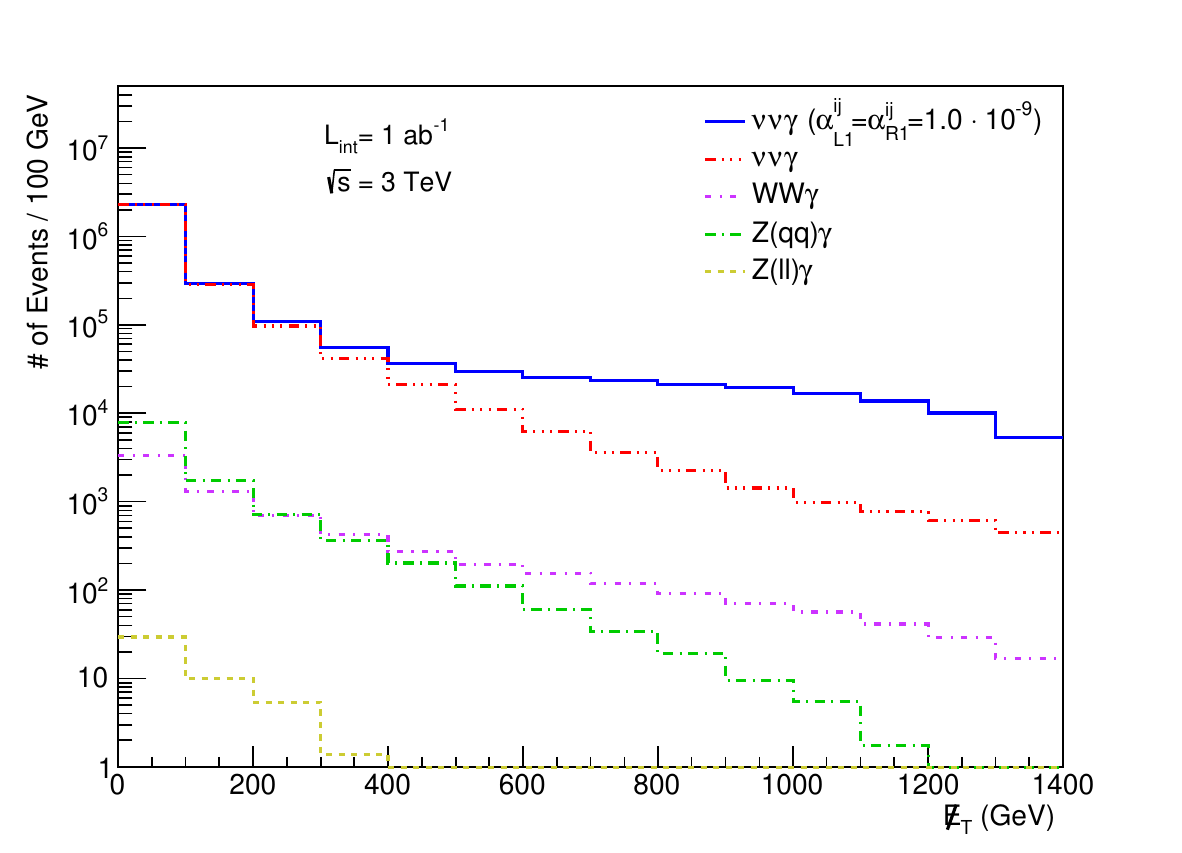}
\includegraphics[scale=0.4]{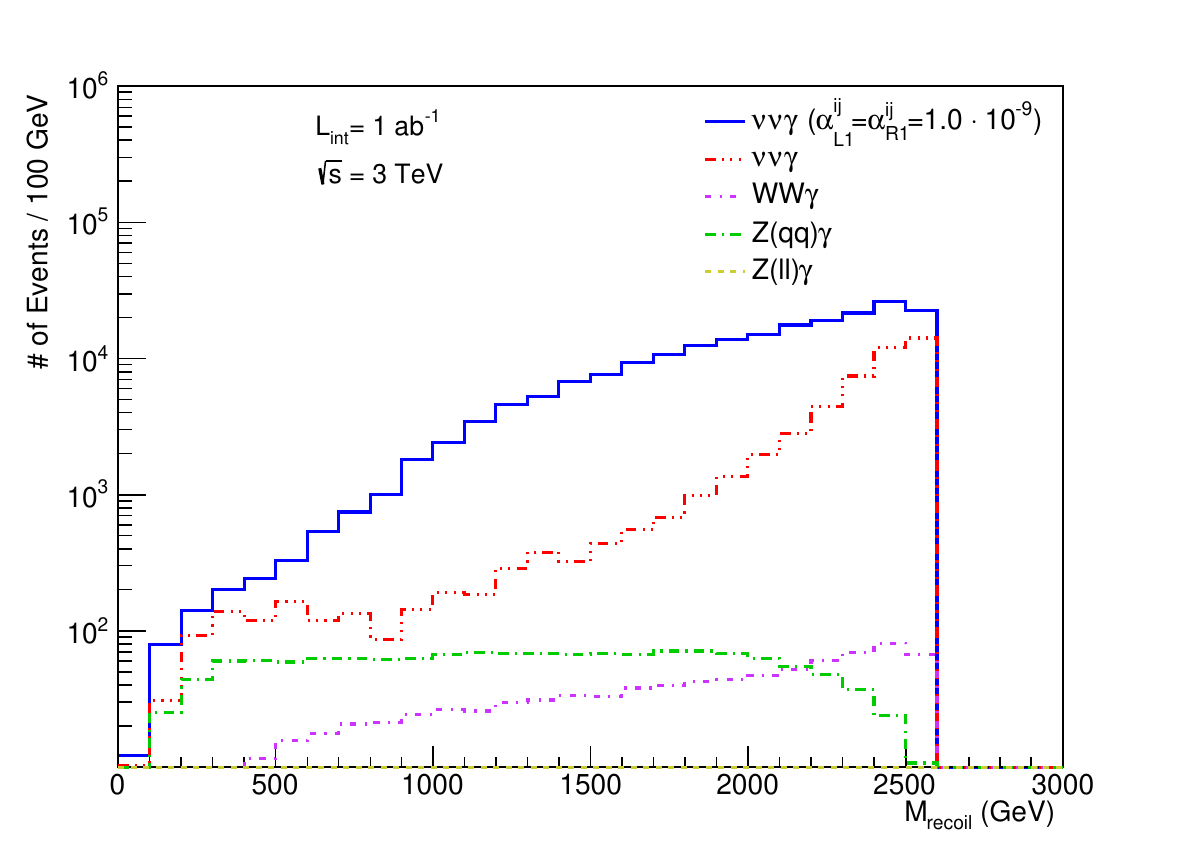}
\caption{The normalized distribution of $p^{\gamma}_T$, $\eta^{\gamma}$, $\slashed{E}_T$ and
$M_{recoil}$ for the $\mu^{+}\mu^{-} \to \nu\bar{\nu}\gamma$ signal and backgrounds at the Muon Collider with $\sqrt{s}=3$ TeV with $L_{int}=1$ ab$^{-1}$.}
\label{Fig.muon_kinematics}
\end{figure}
In Fig. \ref{Fig.muon_kinematics}, we display the normalized distribution of $p^{\gamma}_T$ and $\eta^{\gamma}$ (in the first row) as well as $\slashed{E}_T$ and
$M_{recoil}$ (in the second row) for the signal with $\alpha_{L1}^{ij}=\alpha_{R1}^{ij}=1.0\times10^{-9}$ and relevant background processes at the Muon Collider with $\sqrt{s}=3$ TeV with $L_{int}=1$ ab$^{-1}$  after pre-selection cuts.
After pre-selection, we impose the following further cuts: the final state photon must satisfy the transverse momentum cut $p_T^{\gamma} >$ 400 GeV, the missing transverse energy cut  $\slashed{E}_T >$ 200 GeV and the photon recoil mass cut $M_{recoil} > 110$ GeV. These cut values are set to be as inclusive as possible for an analysis of Muon colliders with the center-of-mass energies of 3 and 10 TeV. As shown in Fig. \ref{Fig.muon_costheta}, normalized $\cos\theta^{\gamma}$ distribution for the
$\mu^{+}\mu^{-}\to\nu\bar{\nu}\gamma$ signal process differ from the relevant SM backgrounds.
\begin{figure}[htb!]
\includegraphics[scale=0.4]{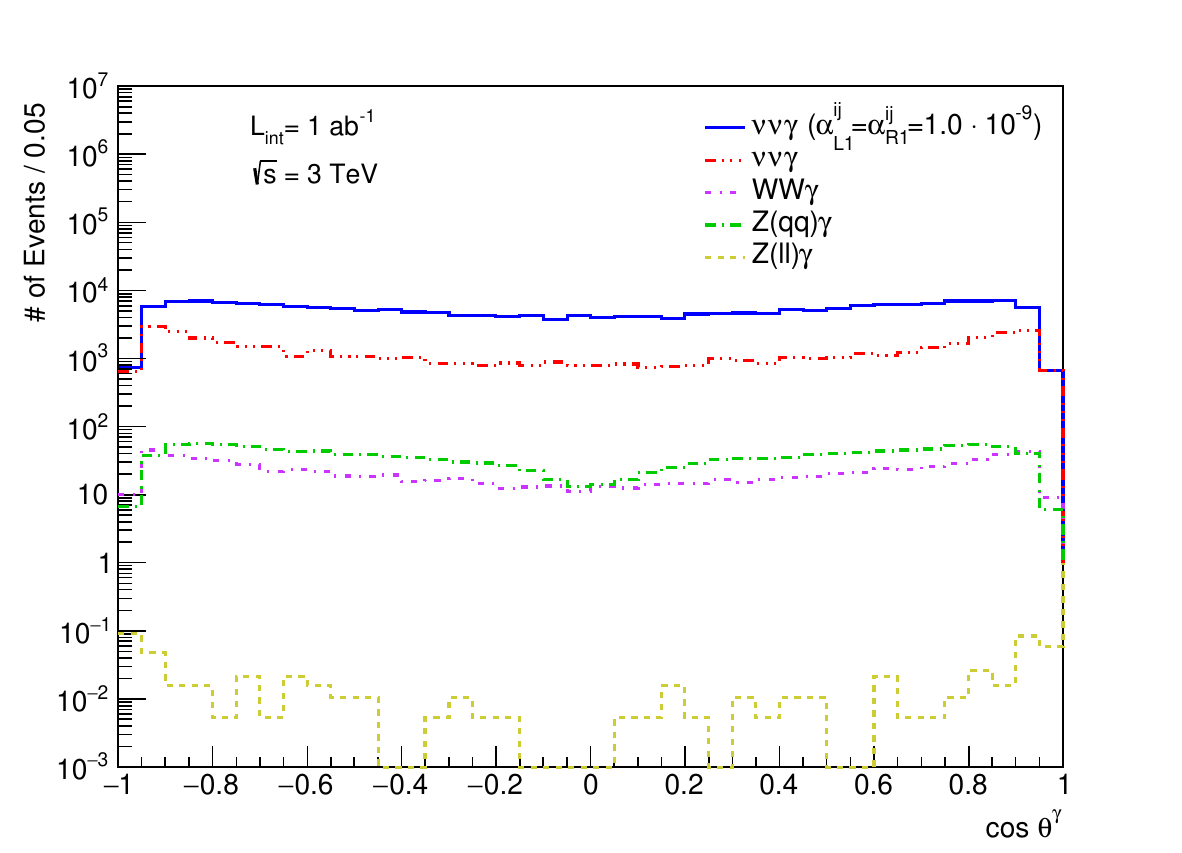}
\includegraphics[scale=0.4]{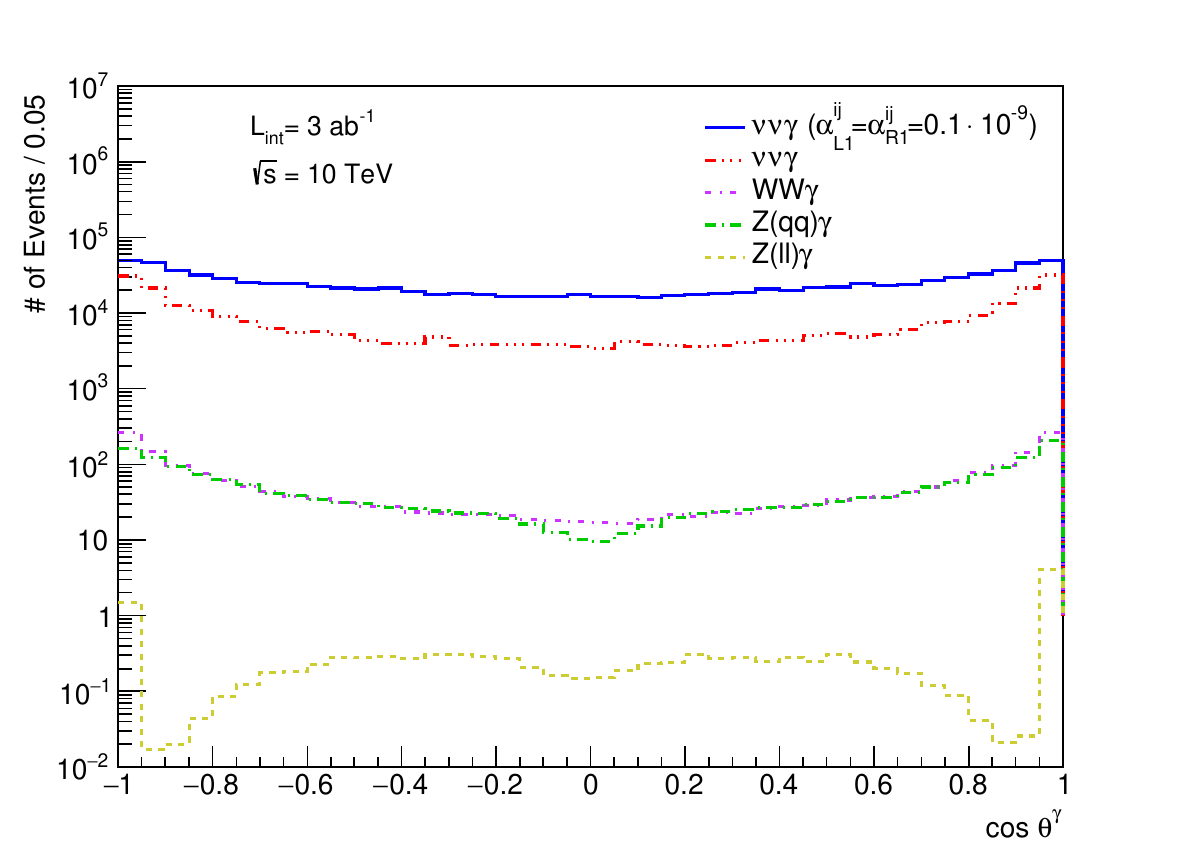}
\caption{The normalized distributions of the polar angle of the photon in the final states for the $\mu^{+}\mu^{-} \to \nu\bar{\nu}\gamma$ signal and relevant SM backgrounds for the Muon Colliders at $\sqrt{s}=3$ TeV on the left panel and $\sqrt{s}=10$ TeV on the right panel.}
\label{Fig.muon_costheta}
\end{figure}
\section{Sensitivity of non-standard neutrino-two photon Couplings}
\begin{figure}[htb!]
\includegraphics[scale=0.4]{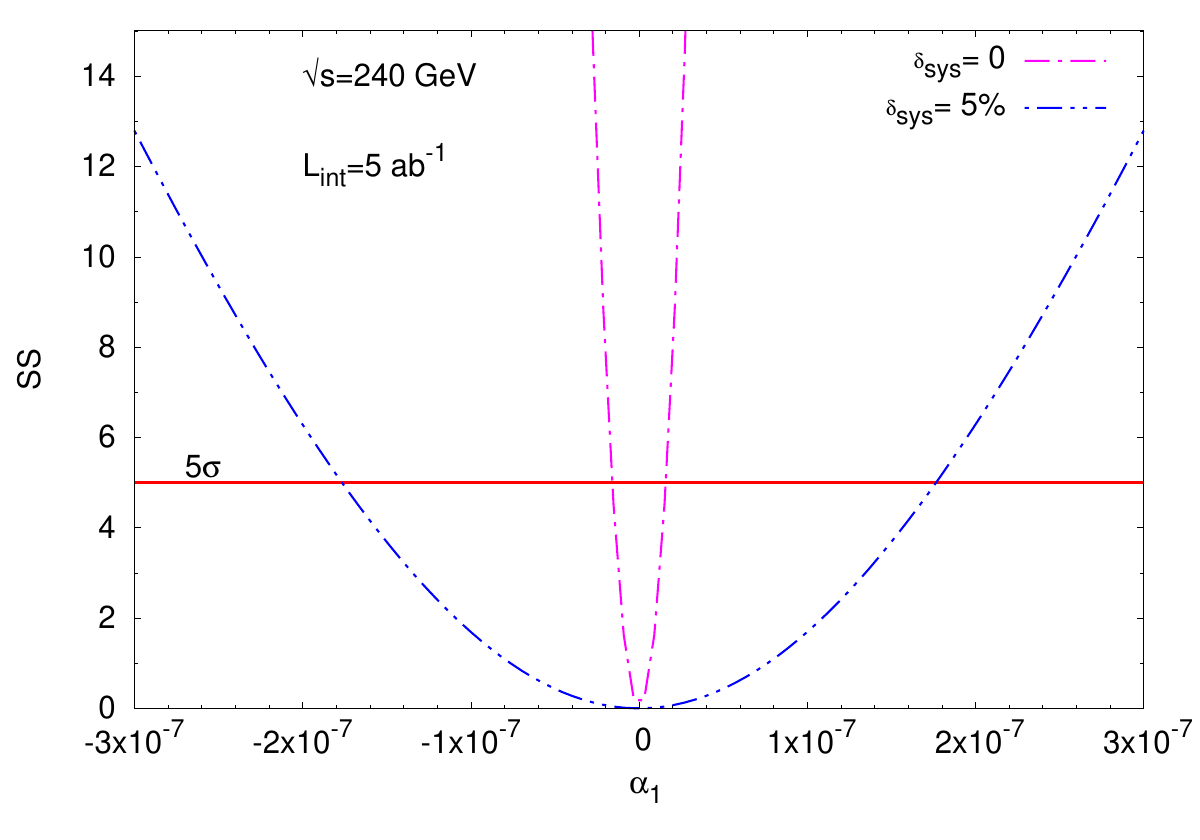}
\includegraphics[scale=0.4]{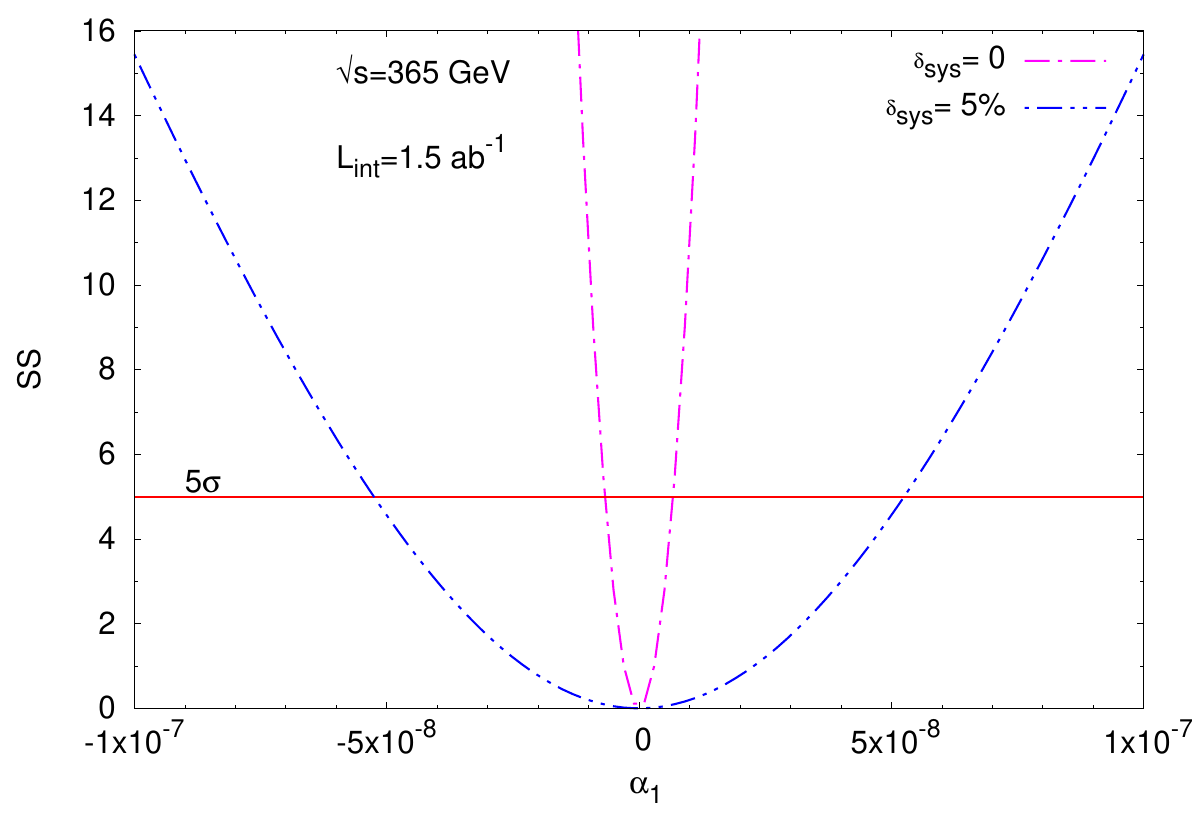}
\caption{The $\mathcal{SS}$ as a function of the $\alpha_1$ coupling without and with $\delta \to 5\%$ systematic uncertainty at the FCC-ee/CEPC with $\sqrt{s}=240$ GeV (on the left panel) and at the FCC-ee with $\sqrt{s}=365$ GeV (on the right panel) with an integrated luminosity L$_{int}$=5 ab$^{-1}$ and 1.5 ab$^{-1}$, respectively.}
\label{Fig.fcc_ss}
\end{figure}

\begin{figure}[htb!]
\includegraphics[scale=0.4]{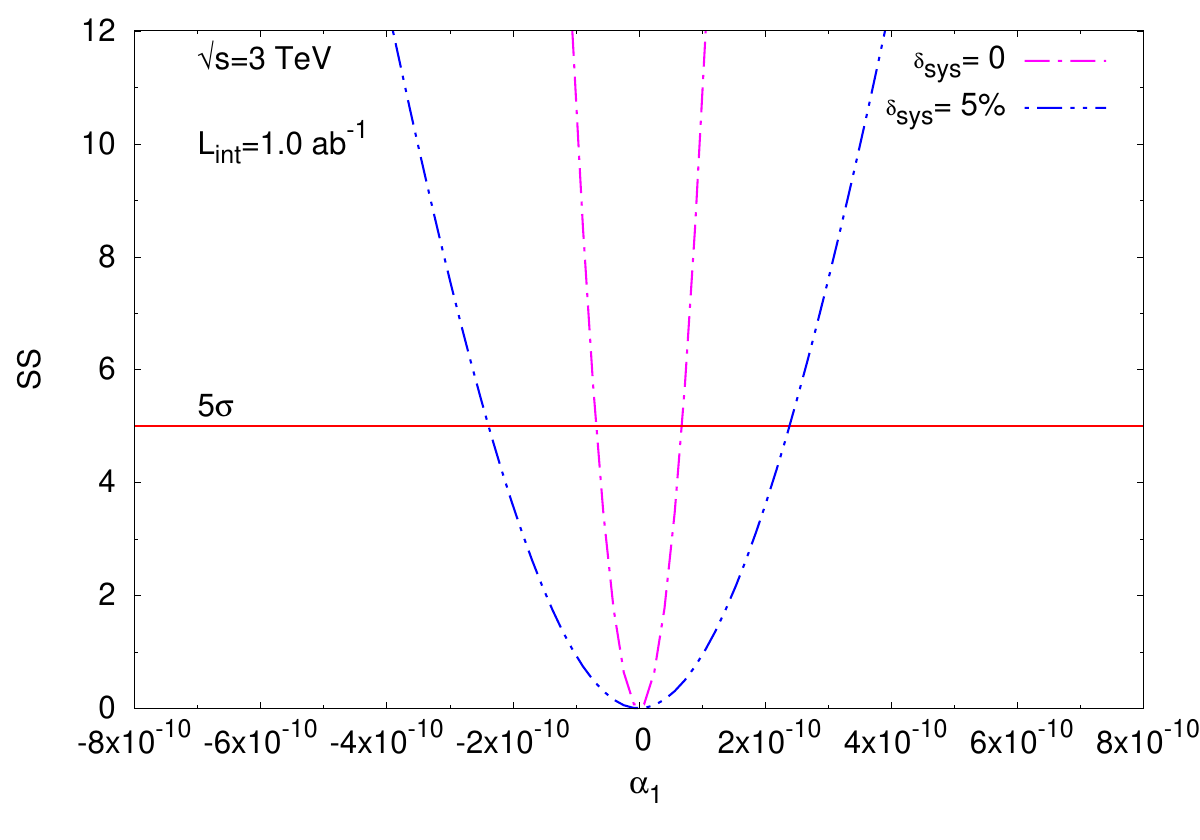}
\includegraphics[scale=0.4]{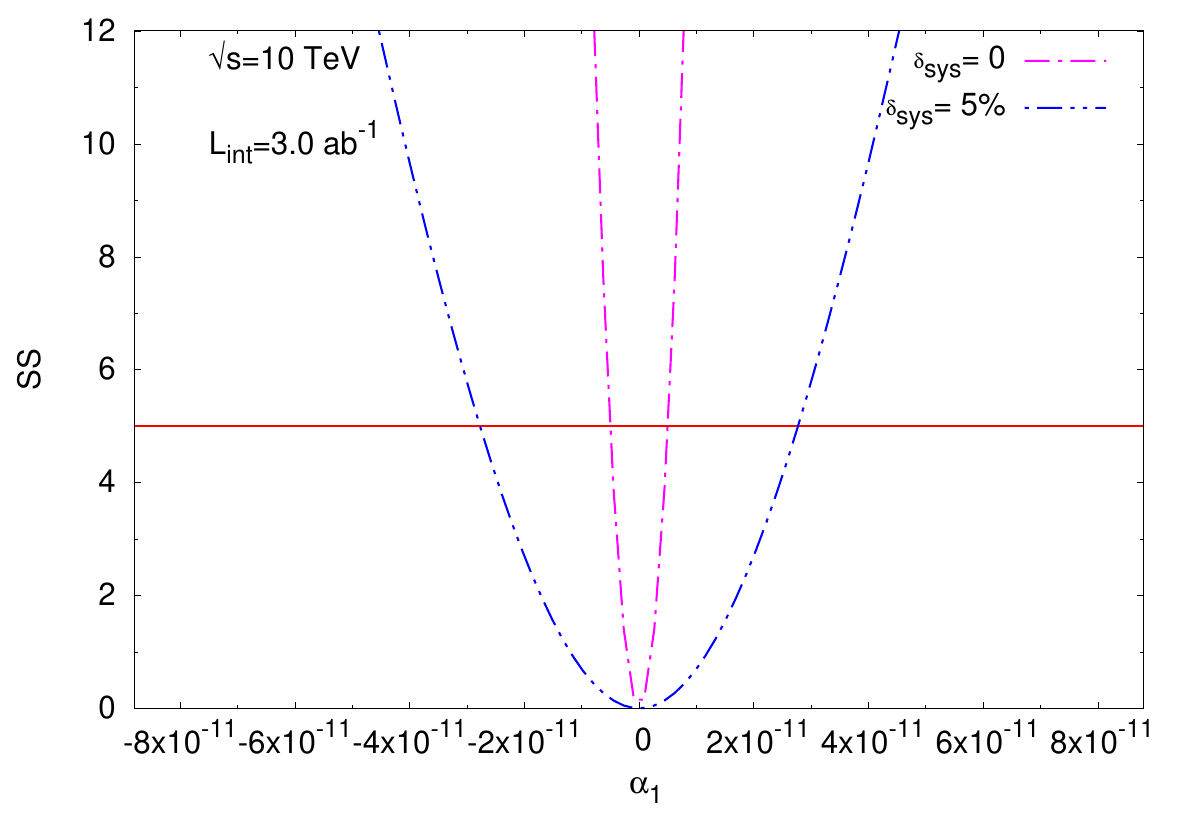}
\caption{The $\mathcal{SS}$ as a function of the $\alpha_1$ without and with $\delta \to 5\%$ systematic uncertainty for muon colliders at $\sqrt{s}=3$ TeV (on the left panel) and $\sqrt{s}=10$ TeV (on the right panel) with an integrated luminosity L$_{int}$=1.0 ab$^{-1}$ and 3.0 ab$^{-1}$, respectively.}
\label{Fig.muon_ss}
\end{figure}

The angular distribution of the final state photon of signal and background processes given in Fig. \ref{Fig.fcc_costheta} and Fig. \ref{Fig.muon_costheta} can be used effectively to find limits on the non-standard $\nu\nu\gamma\gamma$ couplings in $\nu\bar{\nu}\gamma$ production within the Effective Field Theory framework for both $e^{+}e^{-}$ and $\mu^{+}\mu^{-}$ collider options. In order to calculate the median expected statistical significance for $\alpha_1$ coupling, we use the expressions presented in Ref. \cite{Cowan:2010js}:
 
\begin{eqnarray}
\mathcal{SS}&=&
  \sqrt{2\left[(S+B)\ln\left(\frac{(S+B)(1+\delta^2 B)}{B+\delta^2 B(S+B)}\right) -
  \frac{1}{\delta^2 }\ln\left(1+\delta^2\frac{S}{1+\delta^2 B}\right)\right]}
\end{eqnarray}
where $S$ and $B$ are the numbers of events obtained by integrating the normalized distribution of the polar angle of the photon given in  Fig. \ref{Fig.fcc_costheta} and Fig. \ref{Fig.muon_costheta} for the signal and sum of the relevant SM backgrounds, respectively. Also, $\delta$ is the systematic uncertainty. In the limit of $\delta \to 0$, this expression can be simplified as
\begin{eqnarray}
\mathcal{SS} &=& \sqrt{2[(S+B)\ln(1+S/B)-S]}.
\end{eqnarray}
Regions with a $\mathcal{SS}$ $\geqslant$ 5 $\sigma$ are categorized as discoverable regions. Fig. \ref{Fig.fcc_ss} shows the $\mathcal{SS}$ as a function of $\alpha_1$ without and with $\delta \to 5\%$ systematic uncertainty at $\sqrt{s}$= 240 GeV for the FCC-ee/CEPC (on the left panel) and $\sqrt{s}$= 365 GeV (on the right panel) for the FCC-ee options with an integrated luminosity L$_{int}$=5 ab$^{-1}$ and 1.5 ab$^{-1}$, respectively. In Fig. \ref{Fig.muon_ss}, the $\mathcal{SS}$ as a function of $\alpha_1$ without and with $\delta \to 5\%$ systematic uncertainty for the Muon Colliders  at $\sqrt{s}$= 3 TeV (on left panel) and $\sqrt{s}$= 10 TeV (on right panel)  
with an integrated luminosity L$_{int}$=1 ab$^{-1}$ and 3 ab$^{-1}$, respectively. From these figures, limits on $\alpha_1$ can be inferred from the intersection of curves with horizontal red line represented by $5\sigma$ levels. The obtained limits at $5\sigma$ levels on $\alpha_1$ coupling for the FCC-ee/CEPC and the Muon Colliders are given in Table \ref{tab3}.

\begin{table}[htb!]
\caption{The obtained limits at $5\sigma$ levels on $\alpha_1^2$ coupling 
without and with $\delta \to 5\%$ systematic uncertainty
for the FCC-ee / CEPC and the Muon Colliders.}
\begin{ruledtabular}
\begin{tabular}{lcccc}
\multirow{2}{*}{Collider} & \multirow{2}{*}{$\sqrt{s}$ (GeV)} & \multirow{2}{*}{L$_{int}$(ab$^{-1}$)} & \multicolumn{2}{c}{$\alpha_1^2$} \\
         &                  &                      & $\delta \to 0$& $\delta \to 5\%$\\ \hline
FCC-ee / CEPC &240& 5.0  & [-2.56;2.56]$\times 10^{-16}$ &[-3.10;3.10]$\times 10^{-14}$\\
FCC-ee &365 & 1.5 &[-4.57;4.57]$\times 10^{-17}$ &[-2.77;2.77]$\times 10^{-15}$\\
Muon Collider &3000 & 1.5 &[-4.62;4.62]$\times 10^{-21}$ &[-5.71;5.71$\times 10^{-20}$\\
Muon Collider & 10000 &  3.0 &[-3.13;3.13]$\times 10^{-24}$& [-7.67;7.67]$\times 10^{-22}$
\end{tabular}
\end{ruledtabular}
\label{tab3}
\end{table}

\section{Conclusions}

The study of neutrino electromagnetic properties is one of the most active research fields in particle physics.
In the SM, neutrinos do not interact with photons. However, the extension of the SM with massive neutrinos yields $\nu\bar{\nu}\gamma\gamma$ interactions via radiative corrections. Despite the fact that extension of the SM causes tiny couplings, there are new physics models beyond the SM that predict relatively large $\nu\bar{\nu}\gamma\gamma$ couplings. These couplings can be investigated by using higher-order operators with the effective Lagrangian method.

Lepton colliders offer several advantages, especially the clean experimental environment and good knowledge of the initial state. These lead to precise measurements of many reactions.  
They are also well-suited for exploring new physics beyond the Standard Model.

Therefore, we examine the potential of the processes $e^{+}e^{-} (\mu^{+}\mu^{-})\to \nu\bar{\nu}\gamma$ to study $\nu\bar{\nu}\gamma\gamma$ couplings at the FCC-ee/CEPC and the Muon Colliders planned to be built in the near future. We can compare limits on $\alpha_1^2$ summarized in Table III with the only available derived experimental LEP limit \cite{9} and the obtained phenomenological limits in the literature \cite{7,8,mur,6}.

Our best limit on $\nu\bar{\nu}\gamma\gamma$ couplings through the process $\mu^{+}\mu^{-}\to \nu\bar{\nu}\gamma$ at the Muon Collider with center-of-mass energy of 10 TeV is better than the limits by Ref. \cite{7,8,mur} and improves up to a factor of $10^{15}$ compared to the derived experimental LEP limit. However, even including a $5\%$ systematic error, our results are much better than the current available bounds in the literature. 
In Table \ref{tab3}, the anomalous couplings without systematic uncertainty are approximately two orders better than $5\%$ systematic uncertainties limits. 
Finally, if we compare the limits obtained from the Muon Colliders with the limits obtained from the FCC-ee/CEPC, it is clear that muon colliders can achieve limits of eight orders of magnitude better than the FCC-ee/CEPC.

In this study, we have investigated the potential of the $e^{+}e^{-} \to \nu\bar{\nu}\gamma$ process at the FCC-ee/CEPC and $\mu^{+}\mu^{-}\to \nu\bar{\nu}\gamma$ process at the Muon Colliders to probe $\nu\bar{\nu}\gamma\gamma$ coupling. We show that these processes have a great potential to examine this anomalous coupling at the future lepton colliders.

\begin{acknowledgments}
A. Senol would also like to thank the theory division of CERN for their hospitality during his visit, where he had valuable discussions related this work.
\end{acknowledgments}

\end{document}